\documentclass[%
 aip,
 amsmath,amssymb,
 reprint,%
]{revtex4-1}

\usepackage{graphicx}
\usepackage{dcolumn}
\usepackage{bm}

\usepackage[utf8]{inputenc}
\usepackage[T1]{fontenc}
\usepackage{mathptmx}
\usepackage{etoolbox}
\usepackage{xcolor}

\makeatletter
\def\@email#1#2{%
 \endgroup
 \patchcmd{\titleblock@produce}
  {\frontmatter@RRAPformat}
  {\frontmatter@RRAPformat{\produce@RRAP{*#1\href{mailto:#2}{#2}}}\frontmatter@RRAPformat}
  {}{}
}%
\makeatother
\begin{document}

\preprint{AIP/123-QED}

\title[]{Determination of optimal experimental conditions for accurate 3D reconstruction of the magnetization vector via XMCD-PEEM}
\author{Miguel A. Cascales Sandoval}
\affiliation{SUPA, School of Physics and Astronomy, University of Glasgow, Glasgow G12 8QQ, UK}
\affiliation{Institute of Applied Physics, TU Wien, Wiedner Hauptstra{\ss}e 8-10, Vienna, 1040, Austria}%

\author{A. Hierro-Rodr{\'i}guez*}%
\email[Corresponding author e-mails: ]{hierroaurelio@uniovi.es
, Sandra.Gomez@cpfs.mpg.de, amalio.fernandez-pacheco@tuwien.ac.at.}
\affiliation{Departamento de F{\'i}sica, Universidad de Oviedo, 33007, Oviedo, Spain}
\affiliation{CINN (CSIC-Universidad de Oviedo), 33940, El Entrego, Spain}
\affiliation{SUPA, School of Physics and Astronomy, University of Glasgow, Glasgow G12 8QQ, UK}

\author{S. Ruiz-G{\'o}mez*}
\affiliation{Max Planck Institute for Chemical Physics of Solids, 01187 Dresden, Germany}

\author{L. Skoric}
\affiliation{University of Cambridge, Cambridge CB3 0HE, UK}

\author{C. Donnelly}
\affiliation{Max Planck Institute for Chemical Physics of Solids, 01187 Dresden, Germany}

\author{M. A. Niño}
\affiliation{ALBA Synchrotron Light Facility, 08290
Cerdanyola del Vall{\'e}s, Spain}

\author{D. McGrouther}
\affiliation{SUPA, School of Physics and Astronomy, University of Glasgow, Glasgow G12 8QQ, UK}

\author{S. McVitie}
\affiliation{SUPA, School of Physics and Astronomy, University of Glasgow, Glasgow G12 8QQ, UK}

\author{S. Flewett}
\affiliation{Instituto de F{\'i}sica, Pontificia Universidad Cat{\'o}lica de Valpara{\'i}so, Avenida Universidad 330, Valpara{\'i}so, Chile}

\author{N. Jaouen}
\affiliation{SOLEIL Synchrotron, L'ormes des
Merisiers, 91192 Gif-Sur-Yvette, Cedex, France}

\author{R. Belkhou}
\affiliation{SOLEIL Synchrotron, L'ormes des
Merisiers, 91192 Gif-Sur-Yvette, Cedex, France}

\author{M. Foerster}
\affiliation{ALBA Synchrotron Light Facility, 08290
Cerdanyola del Vall{\'e}s, Spain}

\author{A. Fern{\'a}ndez-Pacheco*}
\affiliation{SUPA, School of Physics and Astronomy, University of Glasgow, Glasgow G12 8QQ, UK}
\affiliation{Institute of Applied Physics, TU Wien, Wiedner Hauptstra{\ss}e 8-10, Vienna, 1040, Austria}
\affiliation{Instituto de Nanociencia y Materiales de Arag{\'o}n, CSIC-Universidad de Zaragoza, 50009 Zaragoza, Spain}


\begin{abstract}
In this work we present a detailed analysis on the performance of X-ray magnetic circular dichroism photo-emission electron microscopy (XMCD-PEEM) as a tool for vector reconstruction of the magnetization. For this, we choose 360$^{\circ}$ domain wall ring structures which form in a synthetic antiferromagnet as our model to conduct the quantitative analysis. We assess how the quality of the results is affected depending on the number of projections that are involved in the reconstruction process, as well as their angular distribution. For this we develop a self-consistent error metric which allows to estimate the optimum azimutal rotation angular range and number of projections. This work thus poses XMCD-PEEM as a powerful tool for vector imaging of complex 3D magnetic structures.\\

\end{abstract}

\maketitle

\section{Introduction}

The field of nanomagnetism has rapidly evolved over the last few decades, due to significant advances and developments in fabrication and synthesis methods \cite{fernandez2017three}. These improvements enable to fabricate different natured magnetic systems with complex 3D configurations of the magnetization vector, as opposed to the traditional simple mono-domain magnetic devices. The increase in complexity of magnetic systems \cite{vedmedenko20202020,sander20172017} requires the adaptation and development of versatile characterization methods, where high magnetic sensitivity, spatial and temporal resolutions are some of the most important attributes.

Diverse laboratory-based modern characterization techniques are utilized to study the properties of materials via magnetic imaging, such as: magnetic force microscopy (MFM) \cite{kazakova2019frontiers}, the different Lorentz transmission electron microscopy (L-TEM) modes \cite{phatak2016recent,fallon2019quantitative}, electron holography \cite{thomas2008electron}, scanning electron microscopy with polarization analysis (SEMPA) \cite{lucassen2017scanning,unguris20016}, spin-polarized low energy electron microscopy (SPLEEM) \cite{rougemaille2010magnetic,suzuki2010real}, and the techniques which exploit the magneto-optical Kerr effect (MOKE) to perform wide-field \cite{flajvsman2016high,soldatov2017selective,soldatov2017advanced} or scanning microscopy \cite{flajvsman2016high}.

Analogous to MOKE, although in the X-ray regime, synchrotron-based characterization techniques exploit the strong coupling that exists between photons and magnetism. X-rays offer high lateral resolution due to the short wavelengths, as well as element specificity that arises from the need to tune the photon energy to the absorption edge of the element in question. Imaging setups may be divided in two geometries: transmission and electron yield \cite{le2012studying}. Transmission X-ray microscopy (TXM) \cite{fischer1998imaging,blanco2015nanoscale}, scanning transmission X-ray microscopy (STXM) \cite{zimmermann2018origin} and coherent diffractive imaging (CDI) techniques such as ptychography \cite{shi2016soft} and holography \cite{eisebitt2004lensless}, all analyze the X-rays after passing through the magnetic material. Different strategies may be followed for tomographic reconstruction of the 3D magnetization vector \cite{donnelly2020imaging,hierro20183d,donnelly2018tomographic,hierro2020revealing,donnelly2017three}, depending on the geometry and properties of the sample under investigation. This differs from photoemission electron microscopy (PEEM), or electron yield, where X-rays which have interacted with the material under investigation are not directly collected, but rather the photoelectrons emitted as a consequence of such interaction. Due to the short electron mean free path, PEEM is an excellent candidate for investigating very thin structures close to the surface, \textit{e.g.}, the top layers of a multilayer heterostructure.

Previous works have utilized X-ray magnetic circular dichroism PEEM (XMCD-PEEM) to reconstruct the spatially resolved magnetization vector, by combining images taken at different relative X-ray/sample orientations \cite{le2012studying,ruiz2018geometrically,ghidini2022xpeem,scholl2002x,chopdekar2013strain,chmiel2018observation,digernes2020direct}. Here, we perform a detailed investigation on how the quality of the reconstructed 3D magnetization vector changes depending on the number of projections involved, as well as their angular distribution. For this, 360$^{\circ}$ domain wall (DW) ring structures are chosen as the model to perform the reconstruction, given their small size which pushes the microscope's resolution, and the complex winding sense of the magnetization. These textures are found to form in a synthetic antiferromagnet (SAF) multilayer heterostructure which shows Interlayer Dzyaloshinskii-Moriya interactions (IL-DMI) \cite{fernandez2019symmetry}. For further details on their formation refer to \cite{sandoval2023observation}.

In order to carry out this analysis, the algorithm first aligns the different projections with respect to each other, in such a way that they hold the same spatial orientation. Then, a thorough analysis which consists of running the reconstruction algorithm for different combinations of XMCD projections measured at different angles is performed, applying to the resulting magnetization vectors an error metric that quantitatively gives account of the quality of the reconstruction. The resulting analysis allows to optimize the number of different rotation angles, as well as their specific orientation in relation with the desired accuracy of the magnetization vector reconstruction, being thus very useful for the design of time-efficient XMCD-PEEM experiments.

\section{Methods}

\subsection{Experimental set-up}

The SAF layered structure investigated in this work consists of |Si/Ta (4 nm)/Pt (10 nm)/Co (1 nm)/Pt (0.5 nm)/Ru (1 nm)/Pt (0.5 nm)/CoFeB (2 nm)/Pt (2 nm)/Ta (4 nm)| \cite{fernandez2019symmetry}; where the ferromagnetic layers are asymmetric in material and in thickness. The Co layer has dominating out of plane (OOP) anisotropy enhanced by the Pt layers at the interfaces, whereas the CoFeB layer's thickness has been tuned slightly above its spin reorientation transition (SRT), showing moderately low in plane (IP) anisotropy.

\begin{figure}[ht]
\centering
\includegraphics[scale=0.15]{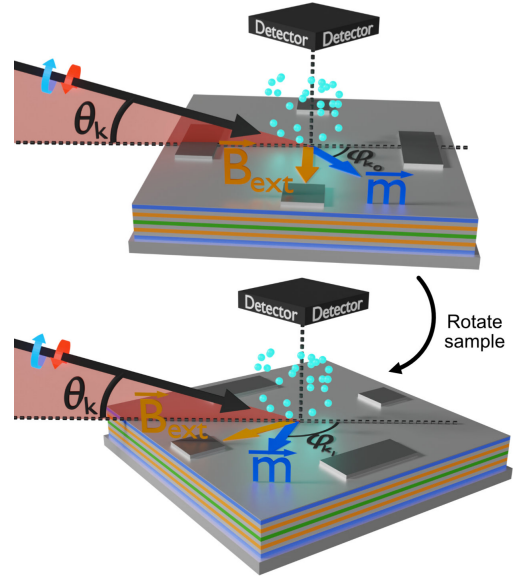}
\caption{\label{fig:fig1} Diagram describing the sample rotation with respect to the X-ray beam for measurement of different XMCD-PEEM projections. The X-ray wave-vector $\Vec{k}$ is given by the black arrow, the circular X-ray polarization eigenmodes by the blue and red circular arrows, the magnetization vector ($\Vec{m}$) by the dark blue arrow, the external magnetic field ($\Vec{B}_{ext}$) by the orange arrow, $\theta_{k}$ is the incidence angle with respect to the surface plane, and $\varphi_{{k}_{0}}$ and $\varphi_{{k}_{1}}$ are the different relative angles between X-ray beam and sample.}
\end{figure}

Prior to performing the synchrotron experiments, a series of repeating $\text{Pt}_{x}\text{C}_{1-x}$ patterns consisting of rectangles and squares were deposited via focused electron beam induced deposition (FEBID) on top of the film surface. Respectively, the size of the squares and rectangles are $1\mu m \times 1\mu m$ and $2\mu m \times 1\mu m$, both being 50 \textit{nm} thick. These are arranged in a square fashion, located at the midpoint of the sides of a $7 \mu m$ square as schematically shown in figure \ref{fig:fig1}. They serve the purpose of providing a non-magnetic signal reference within the field of view (FOV), given that the magnetism dependent photoelectrons do not possess the sufficient energy to escape the sample's surface through this additional bit of material. The non-magnetic signal reference is crucial for properly computing the final XMCD images as there might be slight flux differences and flux spatial distribution when changing polarization, which would alter the amount of emitted photoelectrons inducing ficticious magnetic contrast. Thus, these corrections and references are crucial in order to be quantitative with PEEM.

The microscopy measurements were taken at the PEEM endstation of CIRCE beamline in ALBA Synchrotron \cite{aballe2015alba}. The sample is transferred to the PEEM chamber mounted on a holder with a dipolar electromagnet, providing the capability of applying IP uniaxial magnetic fields \cite{foerster2016custom}. It is mounted in such a way that the nominal easy axis (given by the $\text{Pt}_{x}\text{C}_{1-x}$ rectangle's long axis) is aligned with the external magnetic field direction ($\vec{B}_{ext}$). The system allows rotation of the sample with respect to the surface normal, effectively changing the projection of the incoming X-ray beam onto the sample's directions, as evidenced by figure \ref{fig:fig1}. Measuring at different X-ray/sample relative orientations provides sensitivity to different components of the magnetization vector, given that in XMCD-PEEM magnetic contrast is proportional to $\Vec{k}\cdot\Vec{m}$ \cite{stohr2006magnetism}, with $\Vec{k}$ and $\Vec{m}$ representing respectively the X-ray wave-vector and the magnetization vector.

\subsection{XMCD image measurement and post-processing}

The procedure followed in this work to obtain XMCD images is very similar to the one discussed in \cite{le2012studying}. After reaching the desired magnetic state, 256 images (acquisition time 2s) are recorded for each incoming X-ray circular polarization in order to perform posterior averaging and improve the signal-to-noise ratio. Prior to the subsequent averaging of the same polarization images, a normalization is performed where each individual image is divided in a pixel-wise operation by a largely defocused image in order to remove channelplate contributions. The normalization image is obtained experimentally by going to a homogenous area without obvious features, overfocusing the first lense (about 5$\%$), and taking the average of 64 single images with the same settings as the real data. Once the channelplate contributions are removed, each polarization stack of images is individually aligned in order to correct for potential drifts during the time of measurement. For this, \textit{python's scikit-image} library \cite{van2014scikit} is used, where sub-pixel alignment is performed utilizing its Fourier-space cross-correlation algorithm. The alignment is done by selecting a region of interest (ROI) with a clear, sharp feature, which in this case is chosen to be one of the FEBID deposited landmarks within the FOV. It is crucial to perform the channelplate correction prior to the alignment of each stack, otherwise artifacts due to the translation would be induced. In addition to the image alignment, within each polarization stack, an equalization in image brightness is performed. For this, a proportionality factor that equalizes the average intensity in the $\text{Pt}_{x}\text{C}_{1-x}$ deposits for each image in the stack is found and applied as a global intensity factor to each full image. This is done to take into account and correct for potential X-ray flux variations during the time of measurement.

The averaging of the two aligned stacks of images is now performed giving as a result two averaged images $I_{CL}$ and $I_{CR}$, corresponding respectively to incoming circularly left and right polarization. The cross-correlation algorithm is utilized now again for aligning these two images, and the intensity equalization is similarly done by finding a factor $f$ which relates the intensity in the $\text{Pt}_{x}\text{C}_{1-x}$ deposits, \textit{i.e.}, $f = \overline{I}_{CL}/\overline{I}_{CR}$, with $\overline{I}$ denoting the averaged intensity in the deposits. This factor accounts for changes in signal upon reversing the circular polarization. The final XMCD image is computed as $I_{XMCD} = (I_{CL}-f\cdot I_{CR})/(I_{CL}+f\cdot I_{CR})$ \cite{stohr2006magnetism}, where these are all pixel-wise operations.

\subsection{Magnetization vector reconstruction}

To perform reconstruction of the 3 components of the magnetization vector, a minimum of three different projections are required in order to create a solvable system of equations with unique solutions. Experimentally, this is achieved by rotating the sample in the PEEM chamber about the sample normal and taking XMCD projections at different orientations, as sketched in figure \ref{fig:fig1}. The XMCD images at each of the azimuthal angles are computed utilizing the procedure described in the previous section, although these host different spatial orientations due to the relative rotation between sample and camera. To correct for this, a new protocol which aligns the different azimuthal charge projections (computed as $I_{CL}+f\cdot I_{CR}$) to one another is developed. Charge images are used for this, given that their contrast is independent of the magnetic configuration and azimuthal orientation, unlike the XMCD signals.

First, a single projection's spatial orientation is chosen as a reference, with respect to which the rest of the projections are aligned to. For this, the algorithm finds the most suitable affine transformation parameters: rotation, translation, scale and shear, which take the distorted projection to the reference. Scale and shear adjustments are necessary to correct image deformations introduced by the electron optics upon sample rotation. The error metric defined for this consists of the pixel-wise squared distance between both charge images, and the effectiveness of the procedure is further enhanced by applying a combination of Sobel edge and high-pass filtering algorithms to give more weight to the edges, which serve as alignment features. The optimized affine transformation parameters, which are found from running the algorithm on the charge images, are in the end applied to the corresponding XMCD images.

With the different projections now aligned, the magnetization vector is reconstructed by fitting at each pixel the associated XMCD azimuthal profile to the model, as given by expression \ref{eq:exp1}.

\begin{eqnarray}
\text{XMCD} (\theta_{k},\varphi_{k},|\vec{m}|,\theta_{m},\varphi_{m}) = \vec{k}(\theta_{k},\varphi_{k})\cdot\vec{m}(|\vec{m}|,\theta_{m},\varphi_{m})
\label{eq:exp1}
\end{eqnarray}

$\theta_{k}$ and $\varphi_{k}$ are the independent (or known) parameters which describe the normalized X-ray wave-vector, corresponding respectively to the X-ray incidence angle with respect to the sample's surface, and the azimuthal rotation angle. These angles are known from the experimental setup. The remaining are the unknown (or fit) parameters: $|\vec{m}|, \theta_{m}$ and $\varphi_{m}$, which are the modulus, polar and azimuthal angles of the magnetization vector, respectively. Ten fits are done per pixel, where in each of these different random initial guesses are given to the fit parameters to avoid getting pinned in local minima due to the parameter landscape.

\subsection{Error metric and analysis}

\begin{figure*}
\centering
\includegraphics[scale=0.165]{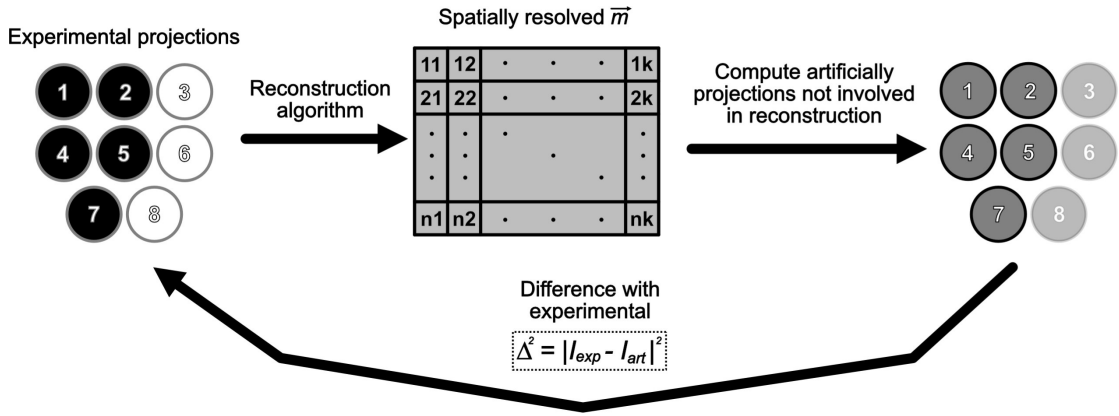}
\caption{\label{fig:fig2} Schematic describing the work-flow of the error metric utilized for quantitatively assessing the quality of the reconstructed magnetization vector. A subset of the initial available experimental projections is taken, in this example 3, 6 and 8 are selected (left white circles). The reconstruction algorithm is applied obtaining the spatially resolved vector given by the matrix, which is then utilized to compute artificially the projections that were not involved in the reconstruction (right, dark gray circles). Finally, these artificially generated projections are substracted in a pixel-wise operation with the experimental ones (black), squaring and summing for all the pixels, and normalizing by the number of images involved (in this particular case 5). This error metric is represented by $\bigtriangleup^2$.}
\end{figure*}

\begin{figure*}
\centering
\includegraphics[scale=0.17]{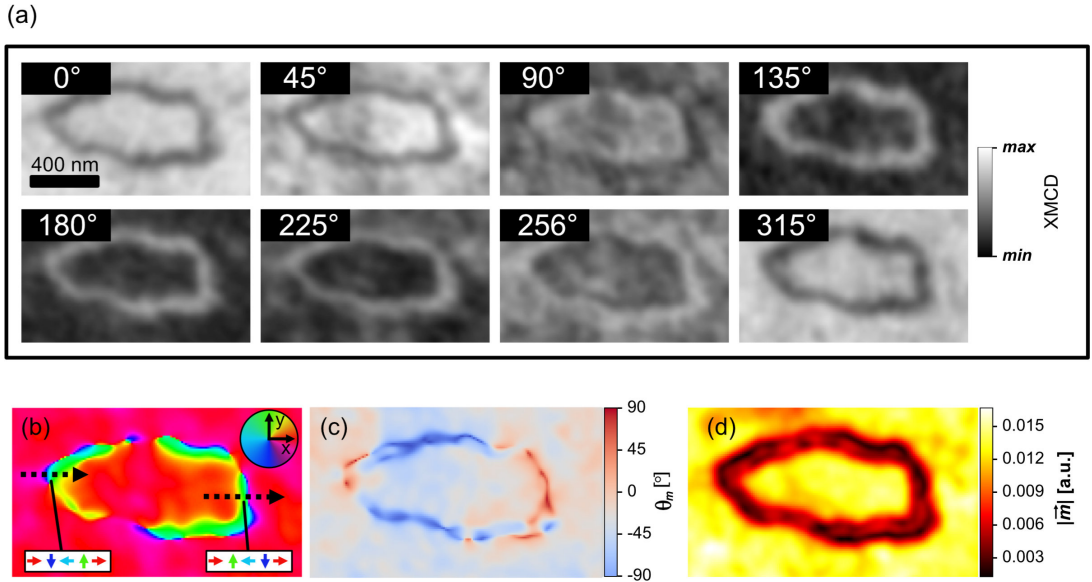}
\caption{\label{fig:fig3} (a) Aligned experimental XMCD-PEEM projections, whose azimuthal rotation angles are given by the numbers in the inset. 0$^{\circ}$ and 90$^{\circ}$ are parallel to the $x$ and $y$ directions of the inset in (b). Spatially resolved (b) IP directions, (c) OOP component angle and (d) modulus of the reconstructed magnetization vector obtained from all 8 experimental projections. The colored arrows in the white boxes of (b) denote the magnetization components along the dashed arrows.}
\end{figure*}

The main objective of this work is to investigate how the quality of the reconstructed results varies depending on the data used, \textit{i.e.}, not only the amount of projections involved, but also if any particular combination of sample orientations are more beneficial than others. In order to be quantitative in this endeavour, an error metric needs to be defined. The procedure followed for this is sketched in figure \ref{fig:fig2}, where 8 is the total number of available projections (since this is the amount measured experimentally). A combination of projections is picked, represented by the white circles (with a minimum of 3 and a maximum of 7), which are then fed to the fitting algorithm to obtain a spatially-resolved magnetization vector. With this vector configuration, the XMCD model is now applied in reverse, artificially generating the projections which were not involved in the reconstruction (black circles of the initial experimental projections). These artificially generated projections are now substracted with their corresponding experimental real XMCD images. The resulting difference images are squared and summed, normalizing the resulting quantity by the number of images involved. The pixelwise error metric corresponding to this process is mathematically described by $\bigtriangleup^2 = |I_{exp}-I_{art}|^2$.

An intuitive way to interpret the meaning of this metric is the following: utilizing part of the available experimental information, the reconstruction algorithm is run. Since the ground truth or real magnetic configuration is not known to compare how accurate the reconstruction is, the only comparison that can be made with real data is with respect to the other experimental projections. In order to do that, these are generated artificially utilizing the XMCD model, and compared in a pixel-wise operation.

\section{Results and discussion}

In previous work, ring-like structures were observed to form within the FOV of the SAF after applying IP demagnetization magnetic field procedures. Briefly, these protocols consist of applying oscillating fields of consecutively decreasing amplitude with a non-zero offset \cite{sandoval2023observation}. To perform vector reconstruction of the magnetization within these rings, 8 projections were measured at the Co $L_{3}$ edge (775.2 eV) with $\theta_{k} = 16^{\circ}$ (large sensitivity to IP components). The signals obtained in this configuration are expected to come exclusively from the top CoFeB layer and not from the bottom Co, as the layered structure prevents the signal from the Co bottom layer to reach the surface due to the short electron mean free path.

The 8 experimental projections are shown in figure \ref{fig:fig3} (a), after having applied the image processing and projection alignment algorithms described in methods. The magnetic signal in these images is determined to be coming mostly from IP components, given that it varies upon azimuthal rotation (OOP magnetization would be insensitive to an azimuthal rotation). The resulting 3D magnetization vector's spherical components obtained after applying the reconstruction fitting algorithm to the 8 projections are shown separately in figures \ref{fig:fig3} (b,c,d). The IP magnetization vector directions, figure \ref{fig:fig3} (b), reveal the presence of 360$^{\circ}$ DW rings separating the outer and inner domains, which point approximately along $+x$. The OOP component, figure \ref{fig:fig3} (c), is very close to zero in the uniformly magnetized areas, although becomes significantly large in the DW area. A large uncertainty is expected for this component, mainly for two reasons. First, the very shallow angle of the incoming X-rays gives small sensitivity to OOP magnetization (proportional to sine of 16$^{\circ}$). Second, in small lengthscales where the magnetization changes rapidly, the resulting magnetic signal measured by the microscope suffers a decrease in amplitude due to the microscope's natural resolution (of the order of 30 nm \cite{aballe2015alba}). Thus even if in reality the signal is coming from IP magnetization, the decrease in amplitude makes the XMCD profile much more susceptible to noise deforming the expected sinusoidal form, and preventing the algorithm from identifying it as such. The decrease in magnetic signal amplitude due to the microscope's resolution is clearly evident in the spatially resolved modulus component, figure \ref{fig:fig3} (d), which becomes significantly smaller in the 360$^{\circ}$ DW (20-30$\%$ relative to the outer uniformly magnetized area). In the ideal case where the microscope had infinite resolution, the modulus of the magnetization vector would be constant throughout the probed space, given that it is made up of the same magnetic material (except if there were inhomogeneities and/or defects which could alter the saturation magnetization). Also, misalignment has a larger negative effect in the quality of the reconstructed results in areas where the magnetic features are of smaller lengthscales, \textit{e.g.}, in the ring.

\begin{figure}[ht]
\centering
\includegraphics[scale=0.22]{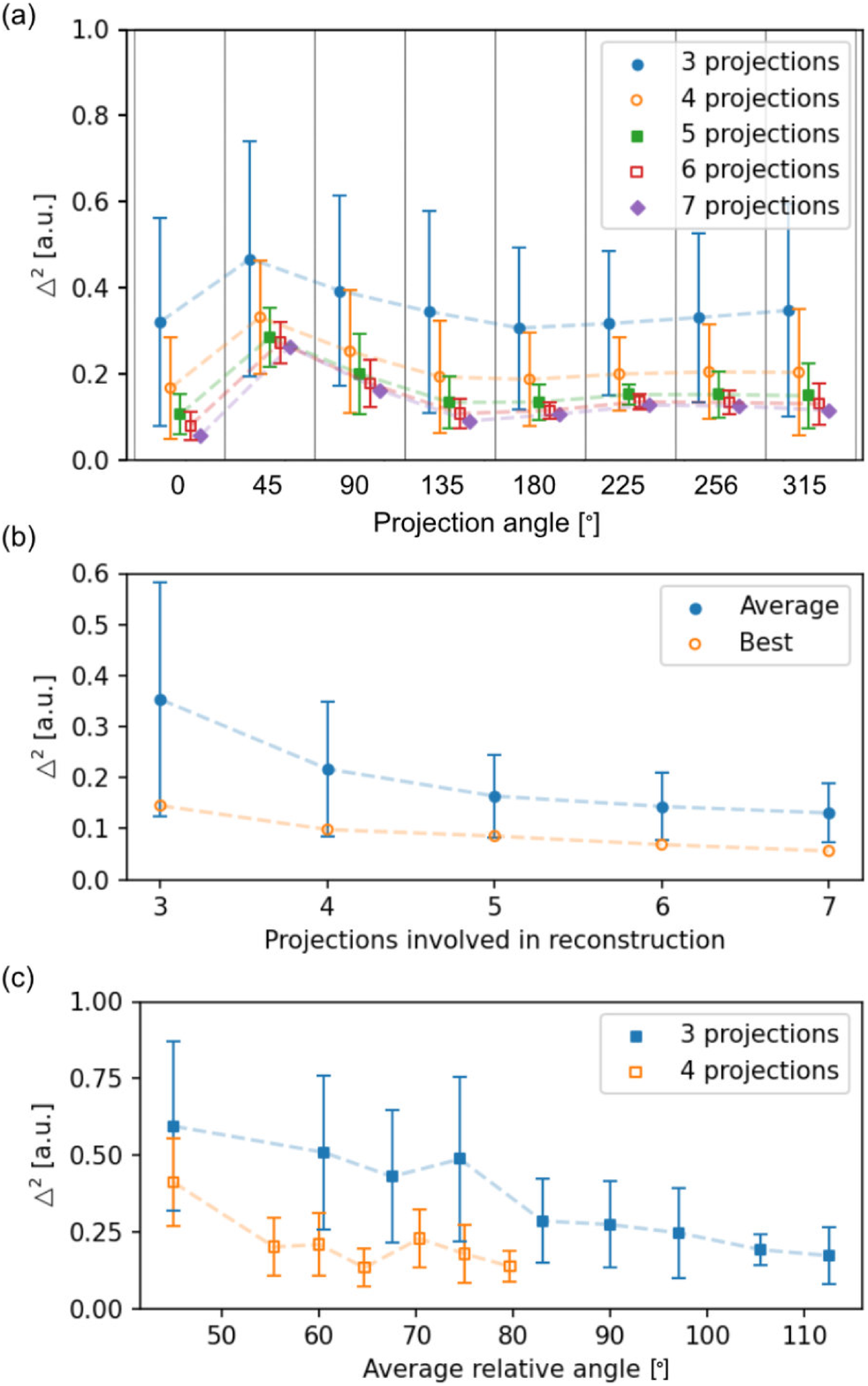}
\caption{\label{fig:fig4} (a) Representation of $\bigtriangleup^2$ for the projections whose azimuthal angle is shown in the x-axis of the plot, for different number of projections involved in the algorithm. (b) Representation of the average (filled circle) and lowest (empty circle) value of the error metric $\bigtriangleup^2$ as a function of the number of projections involved in the reconstruction. (c) Representation of $\bigtriangleup^2$ average with respect the average relative angle in between projections involved for the reconstruction for 3 and 4 projections.}
\end{figure}

The previously described error metric, $\bigtriangleup^2$, is now computed and represented in figure \ref{fig:fig4} (a) as a function of the projection azimuthal angle images displayed on the x-axis. The points in the graph represent the average value of $\bigtriangleup^2$ for all the possible reconstruction combinations which exclude the projection at hand, whereas the error bars give the standard deviation or spread in $\bigtriangleup^2$. This graph gives information regarding the quality of each individual XMCD projection, enabling to identify which of these are reliable, \textit{i.e.}, better levels of signal-to-noise, smaller misalignments and deformations... Overall, the value of the error metric is of the same order of magnitude for all projections, which implies that the noise level and alignment in between the different angles is quite similar, in all showing how the average error decreases as more projections are involved in the reconstruction. A particular case for these experiments concerns the case of the 45$^{\circ}$ projection, where the value of $\bigtriangleup^2$ stands above all, having even a larger error for 7 projections than in the rest of the azimuthal angles with 3. This implies that the image quality at this angle in particular is worse than for the other angles, most probably due to imperfect correction and alignment with respect to the others. This error metric thus allows for detection of bad quality images which can be discarded from the final dataset if needed.

Now, the error metric is plotted with respect to different relevant quantities in figures \ref{fig:fig4} (b,c), as described hereafter. In figure \ref{fig:fig4} (b), the filled circle blue curve shows the averaged error for all the possible reconstruction combinations as a function of the number of projections involved in the algorithm. Differently, the empty circle orange curve represents the smallest error obtained for a single combination of projections, \textit{i.e.}, the best case for each projection number (shown in the supplementary material). In both curves the error decreases as more projections are added to the reconstruction, due to two main factors: increase in statistics, \textit{i.e.}, improving the signal-to-noise ratio, and secondly, from appropriately selecting the azimuthal angles of the projections involved in the reconstruction.

The best case curve is mostly influenced by the increase in statistics, as the optimum azimuthal projection configuration has already been chosen. On the other hand, the blue curve's errors are affected by both statistics and projection angles, as all the possible azimuthal combinations are considered. The effect of selecting the azimuthal projection angles on the quality of the reconstruction is depicted in figure \ref{fig:fig4} (c), where $\bigtriangleup^2$ is represented against the average relative angle in between projections (considering 3 and 4 projections). A very clear trend is observed, which indicates how the error decreases as the spacing between projections becomes larger, converging to similar values for the largest separation possible. This is because the more spread out the projections are, the different components are probed more evenly, having a lower average error in the vector field. Thus, these results reveal, as expected, that it is more effective to have fewer projections evenly spread in space over having numerous projections spanning a narrow angular range. From the best case, 5 projections appear to be a good compromise between quality and time for measurements (each projection takes about 2 hours of measurement including sample rotation). The error for 5 projections improves relatively the best case error obtained with 3 projections by 42$\%$, whereas for 6 and 7 by 52$\%$ and 61$\%$, respectively.

\section{Conclusions}

In conclusion, we quantitatively assess how in XMCD-PEEM, the quality of a reconstructed 3D magnetization vector depends on the number of projections involved and their spatial orientation. For this, we use 360$^{\circ}$ DW ring structures forming in a SAF multilayer as the model to perform a detailed analysis, measuring more than 3 or 4 projections, as is typically done in XMCD-PEEM. We have defined an error metric which uses part of the data for the vector reconstruction, and the remaining for quantitative comparison. As expected, results show how the quality of the reconstructed vector improves significantly upon increasing the number of projections. More importantly, measuring these projections at azimuthal angles evenly spread through the full angular range improve the data quality more efficiently than pure statistics. This quantitative approach provides the reader with an insight for the design of XMCD-PEEM magnetization vector reconstruction synchrotron experiments, where a compromise between accuracy in the reconstruction and duration of the experiments becomes essential.

\section{Acknowledgments}
This work was supported by UKRI through an EPSRC studentship, EP/N509668/1 and EP /R513222/1, the European Community under the Horizon 2020 Program, Contract No. 101001290 (3DNANOMAG), the MCIN with funding from European Union NextGenerationEU (PRTR-C17.I1), and the Aragon Government through the Project Q-MAD. The raw data supporting the findings of this study will be openly available at the Enlighten repository of the University of Glasgow.

A.H.-R. acknowledges the support by Spanish MICIN under grants PID2019-104604RB/AEI/10.13039/501100011033 and PID2022-136784NB, and by Asturias FICYT under grant AYUD/2021/51185 with the support of FEDER funds. S.R-G. acknowledges the financial support of the Alexander von Humboldt foundation. L.S. acknowledges support from the EPSRC Cambridge NanoDTC EP/L015978/1. C.D. acknowledges funding from the Max Planck Society Lise Meitner Excellence Program. The ALBA Synchrotron is funded by the Ministry of Research and Innovation of Spain, by the Generalitat de Catalunya and by European FEDER funds. S.M. acknowledges support from EPSRC project EP/T006811/1. M.A.N and M.F. acknowledge support from MICIN project PID2021-122980OB-C54. For the purpose of open access, the author(s) has applied a Creative Commons Attribution (CC-BY) licence to any Author Accepted Manuscript version arising from this submission.

\bibliography{bibliography}

\end{document}